\documentclass[aps,prb,twocolumn,showpacs,amsmath,amssymb,superscriptaddress
]{revtex4-2} 
\usepackage[english]{babel}
\usepackage{amsmath,amssymb,amsfonts} 	
\usepackage{graphicx}
\usepackage[utf8]{inputenc}
\usepackage{bm}
\usepackage{xcolor}
\usepackage[normalem]{ulem}
\usepackage{SIunits}
\usepackage[colorlinks=true,linkcolor=blue,citecolor=blue,urlcolor=blue]{hyperref}
\usepackage{url}
\usepackage{textcomp}
\usepackage{lineno,hyperref}
\usepackage{float}
\usepackage{chemfig}
\usepackage{rotating}
\usepackage{subcaption}
\usepackage{color} 
\definecolor{red}{rgb}{1.,0.,0.}
\usepackage{soul}
\usepackage{amsmath}

\usepackage{graphicx,wrapfig,lipsum}
\usepackage{caption}
\DeclareUnicodeCharacter{00A0}{~}
\makeatletter
\newcommand*{\rom}[1]{\expandafter\@slowromancap\romannumeral #1@}
\makeatother

\usepackage[inline]{asymptote}

\begin{asydef}
// Global Asymptote definitions
real linkLen=1, linkWidth=2pt;
real rl=2+linkLen;              // distance between beads
guide link=(1,0)--(1+linkLen,0);   // a link
pen beadColor=orange;
pen linkColor=beadColor;
void bead(transform t){
  draw(t*link,linkColor+linkWidth);
  radialshade(t*unitcircle,
    beadColor,shift(t)*(-0.4,0.3),0.01
   ,black,shift(t)*(-0.4,0.3),1.5);
}
pair operator>(pair pos=(0,0), real phi){
  transform t=shift(pos)*rotate(phi);
  bead(t); // draw a bead with a link
  pos+=rl*(Cos(phi),Sin(phi)); // Sin, Cos - in degrees, sin, cos - in radians
  return pos;
};
pair pos;
\end{asydef}
\usepackage{tikz}
\usepackage[labelfont=bf,
justification=raggedright]{caption}
\newcommand\THEOSMARVEL{Theory and Simulation of Materials (THEOS) and National Centre for Computational Design and Discovery of Novel Materials (MARVEL), {\'E}cole Polytechnique F{\'e}d{\'e}rale de Lausanne, 1015 Lausanne, Switzerland}
\newcommand\PSI{Laboratory for Materials Simulations, Paul Scherrer Institut, 5232 Villigen PSI, Switzerland}

\newcommand\Montpellier{Laboratoire Charles Coulomb (L2C), Université de Montpellier, CNRS, Montpellier, France}

\graphicspath{{./images_paper/}}
\begin{document}

\title{Density-functional  perturbation theory for one-dimensional systems: implementation and relevance for phonons and electron-phonon interactions}
\author{Norma Rivano}
\affiliation{\THEOSMARVEL}

\author{Nicola Marzari}
\affiliation{\THEOSMARVEL}
\affiliation{\PSI}

\author{Thibault Sohier}
\affiliation{\Montpellier}

\begin{abstract}
The electronic and vibrational properties and electron-phonon couplings of one-dimensional materials will be key to many prospective applications in nanotechnology.
Dimensionality strongly affects these properties and has to be correctly accounted for in first-principles calculations.
Here we develop and implement a formulation of density-functional and density-functional perturbation theory that is tailored for one-dimensional systems. A key ingredient is the inclusion of a Coulomb cutoff, a reciprocal-space technique designed to correct for the spurious interactions between periodic images in periodic-boundary conditions. This restores the proper one-dimensional open-boundary conditions, letting the true response of the isolated one-dimensional system emerge. 
In addition to total energies, forces and stress tensors, phonons and electron-phonon interactions  are also properly accounted for.
We demonstrate the relevance of the present method on a portfolio of realistic systems: BN atomic  chains, BN armchair nanotubes, and GaAs nanowires.
Notably, we highlight the critical role of the Coulomb cutoff  by studying previously inaccessible polar-optical phonons and Fr\"ohlich electron-phonon couplings. 
We also develop and apply analytical models to support the physical insights derived from the calculations and we discuss their consequences on electronic lifetimes. 
The present work unlocks the possibility to accurately simulate the linear response properties of one-dimensional systems, sheds light on the transition between dimensionalities and paves the way for further studies in several fields, including charge transport, optical coupling and polaritronics.
\end{abstract}

\maketitle

\section{Introduction}
\noindent
 
Over the past three decades, nanostructures have captivated increasing interest, embodying novel physical paradigms and delivering cutting edge technological applications. 
Dimensionality plays the role of an additional degree of freedom,  relevant also beyond fundamental science.  
As regards one-dimensional (1D) systems, our theoretical understanding and the associated first-principles computational tools are yet less developed with respect to the higher-dimensional cases, that is two-dimensional (2D) layers and three-dimensional (3D) bulk materials.
While carbon nanotubes attracted significant attention and success \cite{iijima1991helical,iijima1996structural, thess1996crystalline, journet1997large, kong1998chemical, hone1999thermal}, most first-principles studies have been confined to nanotubes and a limited selection of nanoribbons, thus leaving the important subtleties of the long-range physics associated with these systems and their dimensionality unexplored.

Density-functional perturbation theory (DFPT) is a powerful first-principles tool accurately predicting vibrational properties \cite{gonze1997dynamical,baroni2001phonons,giannozzi1991ab,dal1993ab,PhysRevB.55.10355, sohier2015density}.
In particular, the combination of DFPT along with analytical models has been exploited in the past to reach a comprehensive understanding of phonons, including phenomena such as the well-known LO-TO splitting in 3D systems and its breakdown at the zone center in 2D \cite{cochran1962dielectric,sohier2017breakdown,sohier2016two,sohier2015density}. This approach provides insights into the coupling of these phonons with electrons in both 3D and 2D materials \cite{sohier2016two}.
However, most of the available first-principles codes rely on periodic-boundary conditions in the three spatial dimensions (3D PBCs) and this poses some challenges when dealing with reduced dimensionality.
Indeed, 3D PBCs necessarily imply the simulation of an array of periodically repeated instances of the low-dimensional system, and those periodic images will interact with each other; an effect that is compounded by the lack of screening across vacuum, or even within the low-dimensional systems.
While increasing the amount of vacuum within the simulation cell may suppress  the effects of those spurious interactions for some physical properties, many other properties will always be affected to some extent \cite{sohier2016two, sohier2017breakdown,rozzi2006exact,PhysRevX.9.021050,castro2009exact,PhysRevLett.96.166801,kozinsky2007dielectric, PhysRevX.4.031044}.
This is the case for polar (i.e., with spontaneous net macroscopic polarization \cite{resta1994macroscopic}) or charged and doped systems. 
More generally, this is relevant when long-wavelength perturbations of the charge density are considered.
In all these scenarios, the physical phenomena are indeed driven by long-range (LR) electrostatics which is ultimately ruled by materials dimensionality.
For instance, in linear response, when the electronic charge density of a low-dimensional material is perturbed at momentum $\mathbf{q}$ (in the periodic direction), the reach of the potential generated scales as $\lambda=2\pi/q$ in the non-periodic direction(s). 
Thus, for long-wavelength perturbations (i.e., $\mathbf{q}\to 0$) the spurious interactions persist even for very large distances \cite{sohier2016two, sohier2017breakdown,rozzi2006exact,PhysRevX.9.021050,castro2009exact, kozinsky2007dielectric}. 
In this light, the strategy of increasing the vacuum not only increases the computational cost significantly (linearly/quadratic with the distance in 2D/1D), but also never fully eliminates the issue. 
For momenta smaller than the inverse of the distance between periodic images there will always be the response of a 3D periodic system, instead of the physical 1D one. 
Note that small momenta are exactly those relevant for spectroscopic characterization, charge transport and many other prospective applications.

Proper suppression of these stray interactions across periodic images has been achieved by smoothly truncating the Coulomb interactions between periodic images \cite{rozzi2006exact, castro2009exact}.  
This led to the capability of accounting for materials dimensionality when dealing with excited and neutral properties in 2D \cite{castro2009exact,sohier2015density,pan2018dependence,sun2022evidence,PhysRevLett.108.246803,PhysRevB.87.155304,sohier2021remote,PhysRevMaterials.2.114010,PhysRevB.92.245123}
systems, while only partly in  1D \cite{rozzi2006exact,avraam2012fermi,chan2012capacitance,denk2017probing,pan2018dependence,ping2012ab,prezzi2008optical,rurali2007accurate,wagner2019strain,yang2007quasiparticle,attaccalite2017excitonic,sesti2022anomalous}.
However, computing linear-response properties of  1D materials via DFPT is still an open challenge.
In the following, we address this challenge by developing a DFPT framework tailored for 1D systems.
To this aim, we implement the Coulomb cutoff technique \cite{ismail2006truncation,rozzi2006exact,sohier2015density}  we developed for 2D systems \cite{sohier2015density} within the Quantum ESPRESSO (QE) distribution \cite{giannozzi2009quantum, giannozzi2017advanced} and  we compute consistently potentials, total energy, forces, stresses, phonons and electron-phonon interactions (EPIs).
We also implement the non-analytic contribution to the dynamical matrix to insure smooth  Fourier interpolations and phonon dispersions.
Thanks to those developments, we can highlight the essential role of open-boundary conditions in predicting the correct linear response of 1D systems. 
Namely, we focus the discussion on polar-optical phonons (infrared-active, recently investigated in Ref. \cite{rivano2022infrared}) and Fr\"ohlich couplings, showing for the first time in 1D (to the best of our knowledge) their critical dimensionality signatures. 
Our work is applied to a portfolio of relevant 1D systems including chains, wires and tubes, and the understanding we offer is supported and complemented by analytical models.

The paper is structured as follows.
First, in Section \ref{PBCs_issue} we discuss the challenge posed by PBCs and we illustrate how to rigorously curate this issue by introducing the Coulomb cutoff technique. 
In Section \ref{1D_OBCs}, the implementation of the 1D  framework within QE is detailed.
In Section \ref{results}, we use our developments to study polar-optical phonons (A) and their coupling to electrons (B). We thus discuss the physical understanding provided by our work and, eventually, we comment on the importance of all this for transport applications (C).  
The conclusions follow in Sections \ref{conclusions}.

\section{Inadequacy of 3D PBC\lowercase{s}}\label{PBCs_issue}

First-principles calculations based on plane-wave basis sets rely on PBCs across all three dimensions. When using these PBCs while simulating a system with reduced dimensionality,  periodic images will be present in the non-periodic directions; e.g., nanotubes, nanowires, polymeric and atomic chains.
Our goal consists in isolating the 1D system in such a way that it does not interact with its images, which otherwise would introduce an additional response in our calculations, eventually hindering the true 1D physics.
This is of crucial importance in a variety of cases: systems perturbed at  long wavelengths (even if neutral and non polar),  polar systems, and when doping or charging is included. 
In short, suppressing the stray interactions is essential whenever long-range electrostatics are relevant. 

\begin{figure}[h!]
\centering        \includegraphics[scale=0.38]{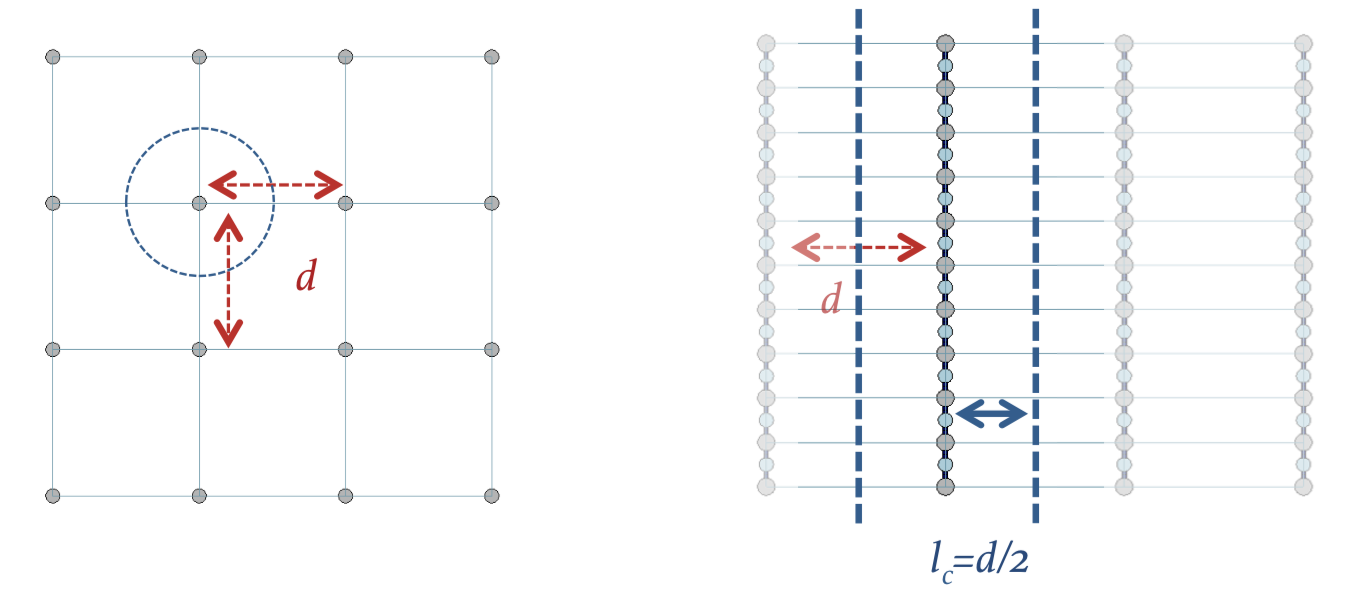}
\captionsetup{width=0.5\textwidth}\caption{Sketch of the supercell construction for 1D systems and the effect of introducing the Coulomb cutoff; after truncation, a given charge in the 1D system interacts only with charges (electrons and nuclei) within a cylinder of radius $l_c$ built around it.}
    \label{sketch_model_pbc}
\end{figure}

Let us start by framing the main concepts and the nomenclature. A 1D system is described as a crystal with periodicity only along  one direction -- $\mathbf{\hat z}$ in this case-- termed \lq in-chain\rq, while having a limited extension (in the range of 1-100 nm) in the the two other \lq out-of-chain\rq\ directions, $\mathbf{\hat x}$ and $\mathbf{\hat y}$. 
Henceforth, the term \lq chain\rq\ might be used to denote a generic 1D system.
The cells in the crystal are identified by $\mathbf{R_z}=m\mathbf{b}$ where $m$ is an integer and  $\mathbf{b}$ is the in-chain primitive lattice vector; the out-of chain components are instead constants. 
The position of each atom $a$ within a cell is then given by $\mathbf{d_a}$ which may contain out-of-chain components depending on the structure (e.g., linear or zig-zag chains, tubes, wires).  
Switching to the reciprocal space, the crystal is described by the reciprocal vector $\mathbf{G_z}$ generated by the in-chain primitive reciprocal lattice vector $\mathbf{b_3}^*$.
Within DFT, the ground state properties of our system are fully determined by the charge density 
\begin{equation}
\rho(\mathbf{r_{\perp}},z)=2e\sum_{\mathbf{k},s}f(\epsilon_{\mathbf{k},s})|\psi_{\mathbf{k},s}(\mathbf{r_{\perp}},z)|^2\,,
\label{density}
\end{equation}
where we sum over the spin-degenerate electronic states, labeled by the in-chain momentum $k_z$ and the band index $s$, $f(\epsilon_{\mathbf{k}, s})$ is the Fermi occupation and 
$\psi_{\mathbf{k},s}$ are the Bloch wavefunctions. The Kohn-Sham (KS) potential, $V_\mathrm{KS}$, for a  neutral/undoped 1D system consists in the external potential created by the ions, i.e., $V_\mathrm{ext}\equiv V_\mathrm{ion}$ in this context,  plus two electronic contributions: the Hartree  $V_\mathrm{H}$ and the exchange-correlation $V_\mathrm{xc}$ potentials. The total potential reads:

\begin{equation}
    V_\mathrm{KS}^\mathrm{1D}(\mathbf{r_{\perp}},z)= V_\mathrm{ext}^\mathrm{1D}(\mathbf{r_{\perp}},z)+
     V_\mathrm{H}^\mathrm{1D}(\mathbf{r_{\perp}},z)+
  V_\mathrm{xc}^\mathrm{1D}(\mathbf{r_{\perp}},z)\,.
\label{potentials}\end{equation}
Each of these potentials has the periodicity of the crystal , i.e., $V(\mathbf{r_{\perp}},z+R_z)=V(\mathbf{r_{\perp}},z)$; the same holds for the electronic density in Eq. \ref{density}. 
It is important to recall here that materials properties can be derived starting from space integrals of the electronic charge density times these potentials.
When using 3D PBCs, rather than simulating the isolated 1D system, one deals with an array of copies obtained by periodically repeating the system in the three dimensions of space with a given amount of vacuum to separate them;
this is shown in Fig. \ref{sketch_model_pbc}. 
Thus, the total potentials from each system, given in Eq.\ref{potentials}, combine as
\begin{equation}
V_\mathrm{KS}^\mathrm{3D}(\mathbf{r_{\perp}},z)= \sum_{\mathbf{R}_{\perp}}V_\mathrm{KS}^\mathrm{1D}(\mathbf{r_{\perp}}-\mathbf{R}_{\perp},z) \,,
\end{equation}
where $|\mathbf{R}_{\perp}|$ describes a square periodic lattice of parameter $d$ in the out-of-chain direction (Fig.\ref{sketch_model_pbc}; we are assuming here a tetragonal cell as a primitive unit of repetition).
The resulting potential $V_\mathrm{KS}^\mathrm{3D} \neq V_\mathrm{KS}^\mathrm{1D}$ satisfies the 3D PBCS.
Then, if the 1D system is perturbed at small momenta $q_z$ (i.e., long-wavelengths) its electronic charge density, periodic only along the z-direction, will respond by generating a potential that is a decaying function of $q_z|\mathbf{r_{\perp}}|$  in the out-of chain directions, long-range in real space for $q_z$ approaching 0. This decay function is intricately connected to modified Bessel functions, as explored in more detail in references such as \cite{rivano2022infrared, giuliani2005quantum}.
As soon as the range of these interactions is comparable with the distance $d$ between the periodic repetitions ($q_z d \lesssim 1$), spurious contributions alter the response of the isolated system.
In essence, PBCs constrain us to simulate a 3D crystal, consisting of weakly bounded 1D substructures, rather than the desired isolated 1D system, and its true physical response will be overlaid with those of all periodic copies.
Similar considerations are valid, besides linear-response, in charged, doped or polar materials where even the energetic and forces may be influenced by the presence of the periodic copies. 

This issue can be addressed via the Coulomb cutoff technique, as successfully demonstrated in several works \cite{ismail2006truncation,rozzi2006exact,castro2009exact, sohier2015density}.
In fact, the conventional approach of simply increasing the vacuum between images only reduces the affected portion of the Brillouin zone (BZ) (the region where  $q_z d \lesssim 1$), while the computational cost significantly increases.
For a more systematic and physical solution, we instead enforce a 1D Coulomb cutoff based on the one proposed in Ref.  \cite{rozzi2006exact}, and implement it in the relevant packages (PWScf and PHONONS) of the QE distribution \cite{RevModPhys.73.515, giannozzi2009quantum,giannozzi2017advanced}.
This implementation leads to the correct 1D open boundary conditions (OBCs) for the computation of potentials, total energies, forces and stress tensors, phonons, and EPIs. 

\section{1D open-boundary conditions implementation}\label{1D_OBCs}

The Coulomb cutoff technique consists in explicitly truncating the spurious interactions between periodic images. 
This is done by modifying the Coulomb kernel, rather than directly the potentials.
The kernel $v_c$ is thus redefined as $\bar{v}_c$:
\begin{equation}
   v_c(\mathbf{r})=\frac{1}{\mathbf{r}} \rightarrow \bar{v}_c(\mathbf r)=\frac{\theta(l_c-|\mathbf r_{\perp}|)}{|\mathbf{r}|}\,,
\label{kernel}
\end{equation}
and all the long-range (LR) contributions to the potentials (i.e., the ones affected by the stray fields, associated to the spurious interactions between periodic images, $V_\mathrm{ion}, V_\mathrm{H}$) are then obtained by convolution of this truncated kernel with the electronic charge density in Eq. \ref{density}   \begin{equation}
     \bar{V}(\mathbf{r})=e\int \rho(\mathbf{r}') \bar{v}_c(|\mathbf{r}-\mathbf{r}'|)\,d\mathbf{r}'\,,
 \end{equation}
in such a way that a given charge in the 1D system interacts only with charges within a cylinder of radius $l_c$ built around it (see sketch in Fig. \ref{sketch_model_pbc}).
Note that although the kernel is discontinuous, the final potentials are smooth thanks to the convolution with the charge density.
Eventually, the material is effectively isolated, meaning that there is no physical 3D periodic system anymore; there is instead a 1D periodic system, repeated in the two additional dimensions of space in order to build potentials that mathematically still fulfill 3D PBCs but physically lead to the true 1D response.

\subsection{1D Coulomb cutoff }\label{cutoff_implementation}
In practice, within the code the potentials (or at least their LR part) are generated in reciprocal space. Thus, the truncated kernel from Eq. \ref{kernel} becomes:

\begin{equation}
\begin{split}
\bar{v_c}(\mathbf{G})=\frac{4\pi}{\mathbf{G_{\perp}}^2+{G_z}^2}[1+G_{\perp}l_cJ_1(G_{\perp}l_c)K_0(G_zl_c)+\\
+G_zl_cJ_0(G_{\perp}l_c)K_1(G_zl_c)]\, ,
\end{split}
\label{kernel_reciprocal}
\end{equation}

where $J_n(x)$ and $K_n(x)$ are, respectively, the $\mathrm{n}^\mathrm{th}$order ordinary and modified cylindrical Bessel functions and $l_c$ is the cutoff length. 
Note that  for the $G_z=0$ plane the expression in Eq. \ref{kernel_reciprocal} is ill-defined since $K_1(x)$ diverges logarithmically for $x \to 0$. However, we are interested in the total potential given as the sum of the Hartree and ionic terms, both modified consistently via this kernel and each defined up to an arbitrary additive constant. Thus, we follow the original development \cite{rozzi2006exact} where this singularity is separated from the truncated kernel and included in these constants.
This is done by considering a cylinder with finite height, $h$, instead of the infinite one of Eq. \ref{kernel_reciprocal}; that is, one first restricts the integration domain along the 1D axis, and then recovers the infinite system by taking the limit for $h \to \infty$. For the purposes of this work, we adopt the same strategy with a crucial variation which will be highlighted in the following. 

We start by defining  $h=N l_0$ as the new length of the finite cylinder, where $l_0$ is a unit-length such that  $h$ is always assumed much larger than the cell size in the periodic direction.
We get the following expression for the Fourier transform of the truncated kernel in Eq. \ref{kernel}:
 \begin{equation}
\bar{v}_c(\mathbf{G_{\perp}}, G_z)=\int_0^{l_c} \int_0^{2\pi}\int_{0}^{h} \frac{e^{-i(G_{\perp}r_{\perp}cos\theta +G_z z)}}{\sqrt{r_{\perp}^2+z^2}}r_{\perp}\, dr_{\perp} d\theta dz\,. 
\end{equation}
Focusing on the $G_z=0$ plane,  we are now left with
\begin{equation}
\bar{v}_c(\mathbf{G_{\perp}}, G_z=0)=\int_0^{l_c} 2\pi J_0(G_{\perp}r_{\perp})\log\Bigg(\frac{h+\sqrt{h^2+r_{\perp}^2}}{r_{\perp}}\Bigg)\,dr_{\perp}\,,
\end{equation}
 where we can substitute $h=Nl_0$ and then split the expression in two integrals, $I_1$ and $I_2$, of which only the first one depends on the height of the cylinder. The truncated kernel now reads:
  \begin{equation} \begin{split}
    \bar{v}_c(\mathbf{G_{\perp}}, G_z=0)= - 2\pi\int_0^{l_c}  J_0(G_{\perp}r_{\perp})\log({r_{\perp}}/l_0)\,dr_{\perp}+ \\
    +2\pi\int_0^{l_c}  J_0(G_{\perp}r_{\perp})\log({N+\sqrt{N^2+(r_{\perp}/l_0)^2}})\,dr_{\perp}\,.
    \end{split}
\end{equation}
The first integral, not dependent on $h$, has the following well-defined solution:
\begin{equation}
    I_1=2\pi \frac{1- J_0(G_{\perp}l_c)-G_{\perp}l_cJ_1(G_{\perp} l_c)\log(l_c/l_0)}{G_{\perp}^2}\,,
\end{equation}
while the second integral, $I_2$,  depends on $h$ and gives:
\begin{equation}
I_2=2\pi l_c \log(2N)\frac{J_1(G_{\perp} l_c)}{G_{\perp}}\,.
\end{equation}
This latter term contains the singularity (i.e., $\lim_{N\rightarrow\infty}I_2= \infty$), which can be dropped by invoking charge neutrality as long as we apply the same cutoff correction to both the Hartree and the ionic potentials. 
At this point, we are interested in fixing the $\mathbf{G}=0$ term of the potential, that is $\bar{V}(\mathbf{G}=0)=0$, adopting a gauge consistent with the existing  3D code and  2D implementation \cite{sohier2015density}.
Thus, we consider the limit of $I_1$ for $G_{\perp} \to 0$ and we get the following behavior:
\begin{equation}
\bar{v}_c(\mathbf{G}_{\perp} \to 0, G_z=0)=\lim_{G_{\perp}\rightarrow 0}I_1 \sim-\frac{\pi}{2} l_c^2[2\log(l_c/l_0)-1]\,.
\end{equation}
The difference in the present approach with respect to Ref. \cite{rozzi2006exact} is the presence of the parameter $l_0$, which comes from the finite height of the auxiliary cylinder.
This parameter has two practical purposes: (1) it enables the use of a dimensionless argument for the logarithm $\log(l_c/l_0)$, and (2) it can be  chosen to set the average potential over the unit cell to zero, i.e., $\bar{V}(\mathbf{G}=0)=\bar{v}_c(\mathbf{G}=0)=0$, leading to:
 $$
 I_1 \sim-\frac{\pi}{2} l_c^2[2\log(l_c/l_0)-1]=0 \to l_0=\frac{l_c}{\exp(0.5)},.
 $$
This corresponds to the conventional choice in QE for both 3D and 2D materials.
 
Eventually, after some manipulations, the final expression for the 1D truncated kernel becomes:
\begin{equation}
 \bar{v}_c(\mathbf{G})=
\begin{cases}
\frac{4\pi}{\mathbf{G_{\perp}}^2+{G_z}^2}[1+G_{\perp}l_cJ_1(G_{\perp}l_c)K_0(G_zl_c)+\\ -G_zl_cJ_0(G_{\perp}l_c)K_1(G_zl_c)]\,,
\hspace{1 cm}
 \mathbf{G_{\perp}},G_z \neq 0\\\
\frac{4\pi}{\mathbf{G_{\perp}}^2}[1-G_{\perp}l_cJ_1(G_{\perp}l_c)\log(\frac{l_c}{l_0})+\\ -J_0(G_{\perp}l_c)]\,,
\hspace{2,5 cm}
 \mathbf{G_{\perp}}\neq 0,G_z=0 \\\
0\hspace{5,5 cm}
\mathbf{G}=0\\
\end{cases}\,.
\label{1D_cutoff_reciprocal}
\end{equation}
Since the $K(x)$ functions damp the oscillations of the $J(x)$ functions very quickly, as pointed out in Ref. \cite{rozzi2006exact}, the cutoff is expected to act only on the smallest values of $G$, while the bulk  behavior ($4\pi/\mathbf{G}^2$) is soon recovered for larger values.
Note that the cutoff $l_c$ needs to be at least as large as the maximum distance between electrons belonging to the system; i.e., the effective thickness of the material $2t$, otherwise some physical interactions of the 1D system with itself will be erroneously cut. In turn, the size of the simulation cell in the non-periodic directions $d$ should be such that electrons belonging to different periodic images are separated by at least $l_c$.
In practice, the cutoff is chosen to be $l_c=d/2$, and the supercell  built such that $d>4t$.

In the following, we detail the implementation of the relevant physical ground-state properties (potentials, energies, forces, stresses) and linear-response ones (phonons and electron-phonon coupling (EPC)).
For the sake of simplicity, we follow the same steps involved in the implementation for 2D systems \cite{sohier2016two}, limiting the present discussion to what is different in 1D with respect to the previous case.

 \begin{figure}
     \centering
    \includegraphics[scale=0.25]{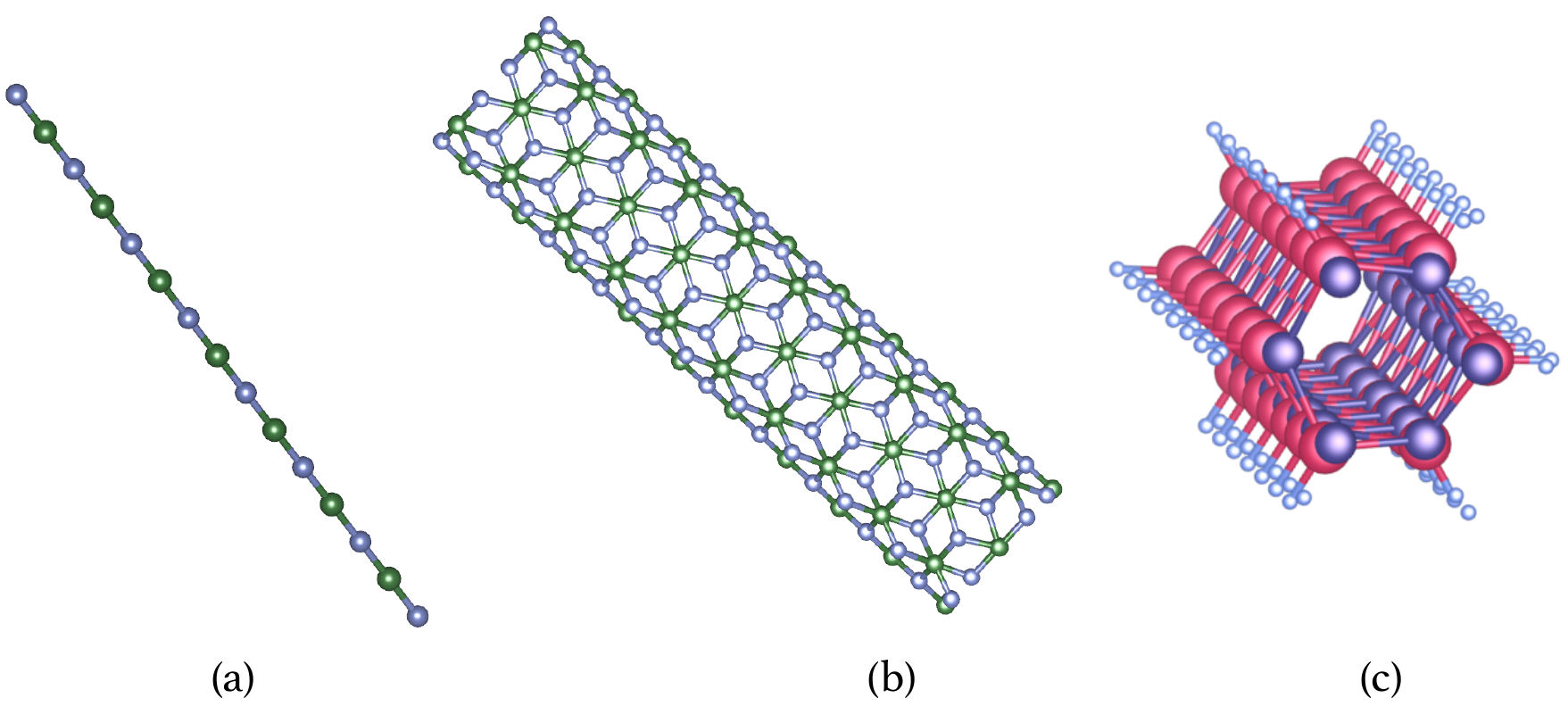}
     \caption{ Crystal structure of (a) a BN atomic-chain, (b) an armchair BN nanotube (4,4), (c) and a GaAs nanowire.}
     \label{portfolio}
 \end{figure}

\subsubsection{Potentials}

The KS potential is the sum  of the external (in this case, ionic), Hartree, and exchange-correlation contributions:
\begin{equation}
    V_\mathrm{KS}(\mathbf{r_{\perp}},z)= V_\mathrm{ext}(\mathbf{r_{\perp}},z)+
     V_\mathrm{H}(\mathbf{r_{\perp}},z)+
      V_\mathrm{XC}(\mathbf{r_{\perp}},z)\,.
\end{equation}
 Here, we are interested in modifying only the LR part of these potentials and thus we can neglect the exchange-correlation term which is short-ranged (SR).
 Note that this  implementation holds for all types of pseudopotentials (i.e., norm-conserving, ultrasoft and projector-augmented wave). 
 Following the conceptual steps of Ref. \cite{sohier2016two}, we proceed by modifying in the QE 3D code the Fourier transform of the local ionic potential and the Hartree potential by substituting the reciprocal expression of the truncated Coulomb kernel.
 This step is straightforward once the kernel has been modified as detailed in the previous section. 
So, we define the local ionic potential as:
\begin{equation}
\begin{split}
    V_\mathrm{ion}^\mathrm{loc}(\mathbf{
    G})=\sum_a e^{-i\mathbf{G}\cdot \mathbf{d}_a}(v_{a}^\mathrm{SR}(\mathbf{G})+v_{a}^\mathrm{LR}(\mathbf{G})) \to \\
    {\bar{V}_\mathrm{ion}^\mathrm{loc}}(\mathbf{
    G})=\sum_a e^{-i\mathbf{G}\cdot \mathbf{d}_a}(v_{a}^\mathrm{SR}(\mathbf{G})+{\bar{v}_{a}^\mathrm{LR}}(\mathbf{G}))    \,,
\end{split}
\end{equation}
and the LR part transforms as
\begin{equation}
\begin{split}
    v_{a}^\mathrm{LR}(\mathbf{G})=-\frac{Z_a}{\Omega}v_c(\mathbf{G})e^{-|\mathbf{G}|^2/4\eta} \to  \\  
    {\bar{v}_{a}^\mathrm{LR}}(\mathbf{G})=-\frac{Z_a}{\Omega}{\bar{v}_c}(\mathbf{G})e^{-|\mathbf{G}|^2/4\eta}
\end{split}
\,,
\end{equation}
where $\Omega$ is the volume of the unit-cell of our 1D system (that is length of the cell times the cross-sectional area in the radial directions).
For the Hartree term we have
\begin{equation}
  V_H(\mathbf{G})=v_c(\mathbf{G})n(\mathbf{G}) \to   \bar{V}_H(\mathbf{G})=\bar{v}_c(\mathbf{G})n(\mathbf{G})\,.
\end{equation}
\begin{figure*}[t!]
\centering
\includegraphics[scale=0.55]{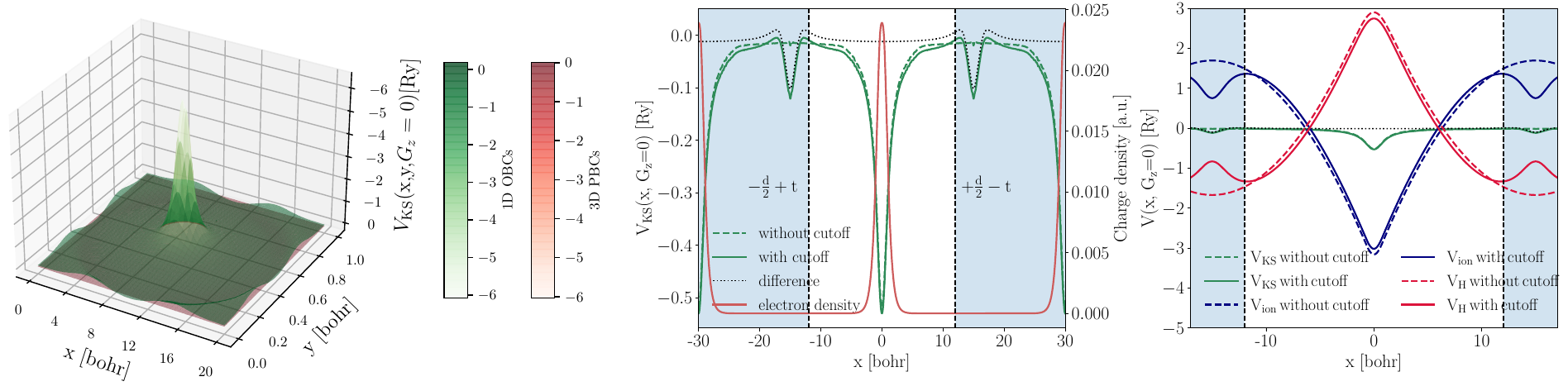}
\caption{Effect of the 1D Coulomb cutoff on the KS potential. In the first panel, we show the KS potential averaged along the chain, with (green) and without (red) the cutoff. 
In the second panel, we focus on the x-axis cross-section of the KS potential in green (dotted/solid for 3D/1D PBCs) and we highlight the physical region; the electronic charge density is reported for reference in orange.
In the last panel, we zoom on the physical region and we plot besides the KS potential also the ionic (blue) and Hartree (red) contributions, always comparing each potential with and without the 1D cutoff (solid and dotted line, respectively). Refer to the main text for details.}
\label{cutoff}
\end{figure*}
As a first validation of the method, we focus on a system with negligible periodic images interactions concerning ground-state properties. This approach allows us to check the modifications introduced so far. We expect the KS potential to be minimally affected by our changes, ensuring the reproduction of the correct physics within the physical region defined by the cutoff length $l_c$. Additionally, we comment on some peculiarities and side effects arising as a consequence of our implementation outside such a region. With this goal in mind, we focus on the simplest system in our portfolio: the BN atomic chain (first panel of Fig. \ref{portfolio}).
We plot in Fig. \ref{cutoff} the total KS potential, as well as its components, without the Coulomb cutoff (3D PBCs), and after its inclusion (1D OBCs).
This is shown in three different panels. 
The first panel offers a three-dimensional representation of the total KS potential averaged along the in-chain direction, $\hat{\mathbf{z}}$, and plotted as a function of the two out-of-chain directions, $\hat{\mathbf{x}}$ and $\hat{\mathbf{y}}$ (which in this case are equivalent by symmetry). 
This representation serves several purposes: (a) it provides insights into the rate of decay of the potential in real-space, (b) aids in detecting any anomalous behavior due to our implementation, and (c) emphasizes the equivalence of the two radial directions. This latter equivalency is crucial for interpreting the subsequent plot. In fact, in the second panel of Fig. \ref{cutoff} we focus on the cross section of the total potential along one of the two equivalent out-of-chain directions.
Together with the KS potential in this case we plot the electronic charge density allowing for a spatial extension comparison. We highlight the limits of the physical region defined by the 1D cutoff by shading the rest of the plot, i.e., for $x>+\frac{d}{2}-t$ or $x<-\frac{d}{2}+t$ with $t$ being the radius of the 1D system (see Sec. \ref{interpol_param}). 
This spatial domain corresponds to the real-space distance over which both the ionic and Hartree potentials exhibit 1D characteristics. Each of these two subsystems, associated with distinct densities, indeed gives rise to an effective cylinder described by the defined cutoff. The intersection between these two cylinders is the one termed the \lq physical region\rq in this context.
Within this physical region, and for this system where neutrality and the absence of out-of-chain dipoles make periodic image interactions weak, $V_\mathrm{KS}$ with and without the cutoff is expected, and found, to be the same up to a constant.
This constant comes from the fact that both KS potentials average to zero, but the one with the cutoff exhibits artifacts, i.e., \lq bumps \rq, outside the physical region, as already discussed for the 2D implementation \cite{sohier2016two}; this is an inevitable consequence of satisfying 3D PBCs.
Finally, in the third panel we zoom on the physical region and we add to the picture also the ionic and Hartree potentials, with and without the 1D cutoff.
Within 3D PBCs, the choice of setting the $\mathbf{G} = 0$ value of the ionic or Hartree potential to zero is equivalent to the inclusion of a compensating jellium background.
At variance with the 2D case, in 1D the potential generated by a linear infinite distribution of charge in the surrounding is logarithmic instead of linear, while in 3D PBC the jellium bath adds a quadratic contribution to the potential between periodic images. 
The correct 1D behavior is restored once the cutoff is applied; however, the effects observed on the cutoff are more subtle with respect to what observed in 2D systems when applying the cutoff \cite{sohier2015density}, at least in the case of neutral non-polar materials considered in this work.
The results depicted in Fig. \ref{cutoff} align with our expectations regarding the impact of the cutoff on both the total potentials and their individual contributions. These findings serve as an initial validation, suggesting that the code is operating as intended. Further validation, as mentioned, would necessitate expanding our investigation to more complex electrostatic systems, such as those with charges or out-of-chain dipoles.

\subsubsection{Energies, forces and stresses}
The total energy per unit cell is computed as
\begin{equation}  E_\mathrm{tot}=E_\mathrm{kin} +E_\mathrm{ext} + E_\mathrm{H} +E_\mathrm{XC}
+ E_\mathrm{i-i}\,,
\end{equation}
i.e., as the sum of the electronic kinetic energy, the energy of the electrons in the external potential created by the ions, the Hartree energy, the exchange-correlation energy and the ion-ion interaction energy. Each of these terms involves a convolution of the charge density with the kernel; thus, we simply need to modify the LR contributions (i.e., $E_\mathrm{ext}$, $E_\mathrm{H}$, and $E_\mathrm{i-i}$) by consistently truncating the Coulomb kernel throughout the code substituting $ v_c(\mathbf{r})$ with $ \bar{v}_c(\mathbf r)$.

Once the 1D potentials and energies are obtained, the forces acting on each ion $a$ are obtained as derivatives of the total energy with respect to the displacement $\mathbf u_{a,i}$ along a given Cartesian direction $i$:

\begin{equation}
 \mathbf{F}_{a,i}=-\frac{\partial E_\mathrm{tot}}{\partial \mathbf{u}_{a,i}}=-\int_{\Omega} n(\mathbf{r})\frac{\partial \bar{V}_\mathrm{ion}}{\partial \mathbf{u}_{a,i}}\,d\mathbf{r}-\frac{\partial E_\mathrm{i-i}}{\partial \mathbf{u}_{a,i}}\,
\end{equation}

where all the terms which do not involve explicitly interactions with the ions have been dropped, and we always imply derivatives at zero displacement $\mathbf{u}_{a,i}=0$.
Thus, we are left with only two terms: the force on an ion coming from the electrons and the contribution given by the interaction with the other ions. Both terms can be obtained as a straightforward consequence of modifying energies and potentials. 

Finally, stresses are computed as derivatives of the total energy  with respect to the strain tensor:
\begin{equation}
    \sigma_{i,j}=-\frac{1}{\Omega}\frac{\partial E_\mathrm{tot}}{\partial \mathbf{\epsilon}_{ij}}\,.
\end{equation}
where $E_\mathrm{tot}$ is proportional to the truncated kernel via the Hartree term (see Ref. \cite{sohier2015density}). 
Here, the LR terms affected by the stray fields are the Hartree term $\sigma_{i,j}^H$ and the contribution coming from the LR part of the local ionic potential, $\sigma_{i,j}^\mathrm{loc,LR}$.
In this case the modifications needed are more extensive and  differ with respect to both the 3D case and the 2D truncation. 
In fact, modifying the kernel and potentials is not enough; one also needs the derivative of the truncated Coulomb kernel with respect to the strain tensor.
By applying the chain rule for the derivative and exploiting the fact that $\partial G_l/\partial \mathbf{\epsilon}_{ij}=\delta_{li}G_j$, we have:
\begin{equation}
\frac{\partial v_c(\mathbf{G})}{\partial \mathbf{\epsilon}_{ij}}= \sum_l
\frac{\partial v_c(\mathbf{G})}{\partial G_l}\frac{\partial G_l}{\partial \mathbf{\epsilon}_{ij}}=
-\frac{\partial v_c(\mathbf{G})}{\partial G_l} 
\end{equation}

In the 1D case we have, based on Eq. \ref{1D_cutoff_reciprocal}:
\begin{equation}
    \frac{\partial \bar{v}_c(\mathbf{G})}{\partial G_z}= -\frac{\bar{v}_c(\mathbf{G})}{\mathbf{G}_p^2+ G_z^2}2G_z\Bigl[1- \beta_z(\mathbf{G}_p, G_z)\Bigr]\,,
\end{equation}
\begin{equation}
    \frac{\partial \bar{v}_c(\mathbf{G})}{\partial |\mathbf{G}_p|}= -\frac{\bar{v}_c(\mathbf{G})}{\mathbf{G}_p^2+ G_z^2}2G_{\perp}\Bigl[1- \beta_p(\mathbf{G}_p, G_z)\Bigr]\,,
\end{equation}
with $\beta_z$ and $\beta_p$ defined as follows:
\begin{equation}
\begin{split}
\beta_p= \frac{|\mathbf{G}|^2}{2G_z \bar{v}_c} \Bigl[-G_{\perp} l_c^2J_1(G_{\perp} l_c)K_1(G_z l_c)- \\l_cJ_0(G_{\perp}l_c)K_1(G_zl_c)+\\ \frac{G_zl_c^2}{2}J_1(G_{\perp}l_c)(K_0(G_zl_c)+K_2(G_zl_c))\Bigr]\,,
\end{split}
\end{equation}

\begin{equation}
\begin{split}
\beta_p=\frac{|\mathbf{G}|^2}{2G_{\perp} \bar{v}_c} \Bigl[\frac{G_{\perp} l_c^2}{2}K_0(G_z l_c)[J_0(G_{\perp}l_c)-J_2(G_{\perp}l_c)]+\\l_cJ_0(G_{\perp}l_c)K_1(G_zl_c)+\\ G_zl_c^2J_1(G_{\perp}l_c)K_1(G_zl_c)\Bigr]\,.
\end{split}
\end{equation}

\subsubsection{Phonons and EPC}
The key ingredient to compute phonon dispersions and EPIs is the response of the electronic density  to a phonon perturbation.
This is obtained in DFPT by solving self-consistently a system of equations in which the unknown is the  lattice periodic part (in italics) of the perturbed KS potential $\frac{\partial\bar{\mathcal{V}}_\mathrm{KS}(\mathbf{r}_p,z)}{\partial \mathbf{u}_{a,i}(\mathbf{q_z})}$. 
In practice, what is needed to compute linear-response properties are the derivatives of the previously defined potentials and energies, already modified, consistently, via the Coulomb cutoff.
Once again we are interested only in the LR terms, i.e., the local ionic $\bar{\textit{V}}^\mathrm{loc}_\mathrm{ion} (\mathbf{q}+ \mathbf{G})$ and Hartree $\Bar{\textit{V}}_\mathrm{H} (\mathbf{q}+ \mathbf{G})$ contributions.
The truncated response is thus obtained by propagating the truncation of these potentials consistently. 
Once this is done, the implementation of the dynamical matrix, from which one obtains the phonon dispersion relations and the EPC matrix elements, is straightforward. The crucial consequence of the cutoff implementation on phonons and EPIs will be at the center of the discussion in the following section. For more details about the implementation of all this in QE, the reader can refer to Ref. \cite{sohier2016two} since the modifications are the same as for the 2D cutoff, just substituting the 1D truncated kernel and potentials.

\subsection{Phonon interpolation and non-analytical corrections}
\label{interpol_param}

Besides the implementation of the 1D Coulomb cutoff, another relevant modification concerns the Fourier interpolation of phonon dispersion relations. Fourier interpolation enables to efficiently compute the full phonon dispersion on a dense momentum grid, first computing the dynamical matrix on a coarse grid, then Fourier transforming it into finite-ranged interatomic force constants (IFCs), and finally Fourier transforming the IFCs back in reciprocal space on a finer momentum grid.
However, in most semiconductors and insulators, non-vanishing Born effective charges (BECs) drive long-ranged  dipole-dipole interactions. These are dimensionality dependent and lead to the IFCs slowly decaying in real-space \cite{piscanec2004kohn,sohier2017breakdown, PhysRevX.9.021050, rivano2022infrared,pick1970microscopic,cochran1962dielectric}. The Fourier interpolation scheme is then not able to fully capture these non-analytic terms since it is based, instead, on the real-space localization of the IFCs \cite{giannozzi1991ab, baroni2001phonons, gonze1994interatomic,gonze1997dynamical,kern1999ab,sohier2017breakdown}.
This prevents from getting accurate phonon interpolations when dealing with polar-optical phonons in a generic $n$-dimensional material.
The standard solution is to build a reciprocal space model for these dipolar interactions and separate the dynamical matrix into a SR and LR component,
\begin{equation}
    D_{a i, b j}(\mathbf{q})= D^{\mathrm{SR}}_{a i, b j}(\mathbf{q})+ D^{\mathrm{LR}}_{a i, b j}(\mathbf{q})
\end{equation}
such that the correct long-ranged contribution to the dynamical matrix can be excluded and then re-added in the interpolation procedure \cite{gonze1997dynamical,sohier2017breakdown, rivano2022infrared}.
This contribution, $D^{\mathrm{LR}}_{a i, b j}(\mathbf{q})$, in 1D has been recently presented in Ref. \cite{rivano2022infrared} and is discussed in Appendix \ref{models}.
It is worth mentioning that the dipole-dipole terms considered here are the leading contribution to the LR IFCs, but higher orders may be present as well with, in general, much smaller consequences on the phonon dispersion relations \cite{PhysRevX.9.021050, ponce2023long}. 
Note that in any case direct phonon calculations (i.e., without interpolation) with the 1D cutoff include all order of multipoles.

The implementation of $D^{\mathrm{LR}}_{a i, b j}(\mathbf{q})$ requires several physical quantities.
Masses, eigenvectors, eigenvalues and BECs are directly obtained from the underlying DFT and DFPT calculations.
What need to be parameterized are instead the effective radius $t$ of the 1D system and its macroscopic dielectric tensor $\epsilon^\mathrm{1D}$, which in this model is replaced by the dielectric component $\epsilon_z$ (i.e., isotropic assumption).
As discussed in the literature \cite{sohier2016two, electric, dielectric, Libbi_2020, PhysRevLett.96.166801, tian2019electronic, andersen2015dielectric, mohn2018dielectric, sohier2021remote, sponza2020proper}, the dielectric tensor concept is ill-defined in nanostructures.
More advanced strategies have been proposed to model the response of two-dimensional materials \cite{andersen2015dielectric, mohn2018dielectric, sohier2021remote, sponza2020proper} based instead on the polarizability and a particularly fundamental and robust theory has recently been proposed in 2D \cite{PhysRevX.9.021050}.
In our work \cite{rivano2022infrared}, however, modelling the material as a dielectric cylinder implies the presence of a 1D dielectric tensor in the analytical model.
This differs from the one computed in PBCs such as in QE, $\epsilon^{\mathrm{QE}}$, which depends on the size of the simulation cell.  
Following the prescriptions from Ref. \cite{rivano2022infrared}, we define:
\begin{equation}   \epsilon_\mathrm{1D}=\frac{c^2}{\pi t^2}(\epsilon_\mathrm{QE}-1)\,,
\end{equation}
where $c$ is the out-of-chain length characterizing the supercell geometry (assumed to be the same in the $\hat{\mathbf{x}}$ and $\hat{\mathbf{y}}$ directions).
In practice, in implementing the correction to the dynamical matrix, we automatize the choice of the effective radius as $t=d/4$, where in the most general case $d$ is the size of the cell in the non periodic direction.
This choice is reasonable assuming that $d$ has been chosen as the minimum size to satisfy the cutoff requirements as explained in Sec. \ref{cutoff_implementation}.

\section{Application to 1D systems}\label{results}
In most practical cases, atoms in semiconductors and insulators carry non vanishing BECs corresponding to a polarization charge inside the material.
This is the origin of non analytic LR contributions affecting phonons (such as polar-optical phonons) and EPIs (e.g., Fr\"{o}hlich and piezoelectric). 
Due to the long-range nature of the phenomenon, dimensionality has profound consequences on linear-response properties. 
The 1D cutoff is thus crucial for their first-principles description.  In this section, we apply our  developments to the following systems: a BN atomic chain, a BN armchair nanotubes of increasing radius, and a GaAs nanowires (see Fig. \ref{portfolio}).
Namely, we show the impact of the present implementation on polar-optical phonons (A) and their coupling to electrons (B), envisioning the related consequences in terms of charge transport (C).

\subsection{Polar-optical phonons}
\begin{figure*}[ht!]
\centering
\includegraphics[scale=0.5]{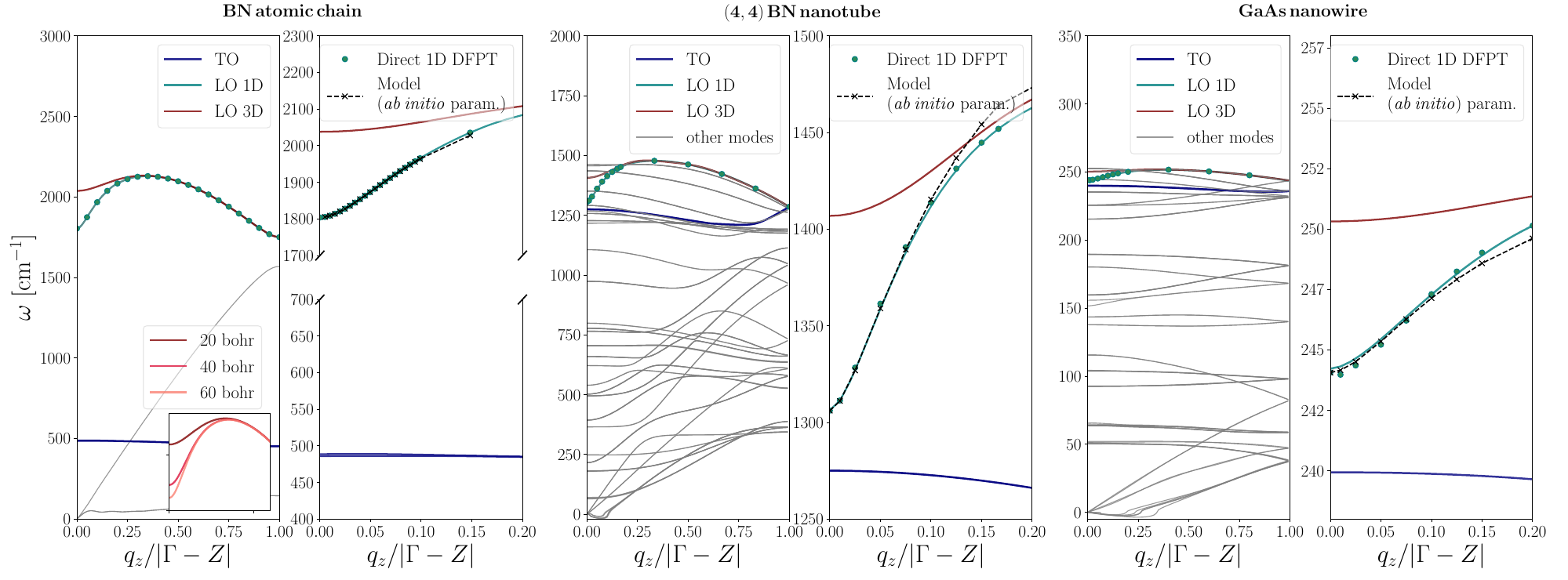}
\caption{Phonon dispersion of a BN atomic-chain, a (4,4) nanotube, and a wurtzite GaAs nanowire with a primitive cell of 24 atoms (including the saturating hydrogens).
For each material, the left panel compares 3D-PBC with 1D-OBC DFPT calculations, explicit for 1D (symbols) and interpolated  for both (lines). The respective right panels show the agreement of our model with the 1D DFPT calculations for the LO branch in the long-wavelength limit.}
\label{BN_atomic-chain}
\end{figure*}
Polar-optical infrared-active phonons --namely the longitudinal optical ones (LO)-- can generate a LR electric field, macroscopic in the long-wavelength limit \cite{born1954k,PhysRev.59.673, pick1970microscopic}
This field is felt by the atoms as an addition to the energy cost associated to their displacement and leads to a blue shift of the phonon frequencies.   
While the strength of this effect is modulated by the dielectric properties of the material (BECs and the high-frequency limit of the dielectric tensor $\epsilon^{\infty}$), its dependency on phonon momenta and size (thickness in 2D, and  radius in 1D) is constrained by dimensionality. 
This phenomenon has undergone extensive investigation in both 3D \cite{pick1970microscopic, cochran1962dielectric} and 2D materials \cite{sohier2017breakdown}. However, its investigation in 1D systems is a relatively recent development \cite{rivano2022infrared}.
In our previous work \cite{rivano2022infrared}, we described polar-optical phonons in 1D systems and presented an analytical model. This model was parameterized using first-principles calculations, allowing for an accurate description of these phonons.
Our findings revealed notable differences compared to 3D systems, where the dielectric shift remains constant across the Brillouin Zone (BZ), commonly referred to as longitudinal-transverse optical splitting. In nanostructures, however, this shift deviates from the 3D behavior. In 2D materials, it exhibits an asymptotic linear relationship, characterized by $\omega \approx q$ \cite{sohier2017breakdown}. In the 1D case, it follows a quadratic-logarithmic relationship as $q^2\log(q)$ \cite{rivano2022infrared}.
In the following we emphasize, based on our previous work, the importance of the boundary conditions and the suitable phonon-interpolation scheme to accurately describe these phonons.

Due to the long-range nature of the interactions they generate, polar-optical phonons are a prime example of dimensionality effects.
The first panel of Fig. \ref{BN_atomic-chain} shows the phonon dispersion for each of the materials considered. 
We compare the results from the standard QE code, i.e., 3D PBCs, and the ones obtained after implementing the 1D Coulomb cutoff technique as explained above, i.e., 1D OBCs. 
Note that the relevant LO and transverse optical (TO) modes, labeled as such in the long-wavelength limit, are highlighted with colors. The other modes are reported in grey. 
For each material, in the left panel of Fig. \ref{BN_atomic-chain} we show the effect of our implementation on the phonon dispersion relations and we highlight that the main difference is found in the long-wavelength limit of the LO branch.
In fact, 3D DFPT predicts a rather flat LO mode (red) with a dielectric shift at $\Gamma$ which  adds up to the mechanical splitting with respect to TO phonons (blue). 
Here, \lq mechanical splitting\rq\ denotes the energy separation LO (in-chain) and TO (out-of-chain) modes, originating not from dielectric effects but rather from inherent distinctions in atomic displacements along the chain or or perpendicular to it.
These features are similar to those commonly found in 3D bulk materials and are related to the presence of the periodic images.
On the contrary, within the current 1D implementation, the dielectric shift experienced by LO phonons (green) is shown to vanish in the proximity of $\Gamma$ with an asymptotic trend in agreement with the expected 1D signature, as discussed in Ref. \cite{rivano2022infrared}.
This is clearly shown in the right panels, where we zoom in on the long-wavelength behavior of the LO mode and we plot the analytical model from Ref. \cite{rivano2022infrared} to support our analysis (see also Appendix \ref{models} for additional details).
Fig. \ref{BN_atomic-chain} also shows that the interpolation scheme proposed here is successful. Thus, the combined effects of the truncated Coulomb interaction and the 1D interpolation scheme enable an accurate description of vibrational properties.
Note that, in nanostructures, phonon calculations at $\Gamma$ do not pose any issue, contrary to the 3D case. Direct phonon calculations give the correct results with or without spurious effects because of the vanishing non-analytic contribution to the zone-center dynamical matrix.\cite{sohier2017breakdown, rivano2022infrared}
Similarly, the short-wavelength limit (i.e., $q_z \to \infty$) is not dependent on dimensionality.
This explains why the correction introduced by the Coulomb cutoff only affects long-wavelength phonons. 

In addition, we emphasize that the discrepancy between the 3D and 1D boundary conditions depends on the vacuum present within the simulation cell, specifically the distance between periodic images in 3D PBCs. The momentum range where the two approaches differ is directly related to it. A larger vacuum leads to a smaller region affected near $\Gamma$ and consequently a softer behavior of the LO branch, asymptotically approaching the 1D limit. This behavior is illustrated in the inset of Fig. \ref{BN_atomic-chain}, where the LO characteristics at small momenta are shown for various distances between periodic images. The 1D  behavior is only  fully recovered when a cutoff is applied. Indeed, regardless of the vacuum size and for sufficiently small momenta, the presence of spurious interactions consistently induces a response resembling that of a 3D periodic system.

\subsection{Fr\"{o}hlich electron-phonon interactions}

Similar to phonons, electron-phonon interactions can undergo significant modifications due to dimensionality.
One prominent example is the Fr\"ohlich coupling between the electrons and the polar-optical phonons discussed in the preceding section.
The long-range nature of this interaction gives rise to distinct signatures in 3D, 2D, and 1D systems.
In 3D, the Fr\"ohlich interaction is known to diverge as $\mathbf{q}\rightarrow 0$ \cite{vogl1976microscopic}, while in 2D, it converges to a finite value \cite{sohier2016two}.
The behavior of the Fr\"ohlich interaction in 1D systems remains unclear, and our implementation can provide valuable insights in this regard.

Here, we focus on the simplest yet instructive system in our portfolio, the BN chain. 
The dispersion relations for small-momentum LO phonons are shown in the second panel of Fig. \ref{BN_atomic-chain}.
We investigate how this mode couples with the electrons in the system by considering phonon-scattering of an electron from an initial state $|\mathbf{k}_i\rangle$ to a final state$|\mathbf{k}_i + \mathbf{q}\rangle$ within a given band $n$ (i.e., intraband scattering only).
Namely, we restrict our analysis to the conduction and valence bands with initial states being at the edge of the BZ, i.e., $Z$ point, that is the conduction (valence) band minimum (maximum). 
\footnote{Note that the electronic bands of the chain exhibit double degeneracy Z for both the valence and conduction bands. To determine the overall coupling, we consider the intraband coupling strength ($g$) for each individual band, sum the interband contributions, and subsequently divide the result by two.}
The EPC matrix elements are defined in DFPT as proportional to the potential perturbation induced by a phonon displacement of atom $a$ in direction $i$ \cite{baroni2001phonons}:
\begin{equation}
g^{\mathrm{DFPT}}_\nu (\mathbf{q}_z)= \sum_{a,i} \frac{\mathbf{e}^{a,i}_\nu (\mathbf{q}_z)}{\sqrt{\hbar/2m_a \omega_\nu (\mathbf{q}_z)}} \langle \mathbf{k}_i +\mathbf{q}| \Delta^{a,i}_{\mathbf{q}_z} \mathcal{V_\mathrm{KS}(\mathbf{r})}|\mathbf{k}_i \rangle \,,
\label{el-ph_coupling}
\end{equation}
where $\mathcal{V_\mathrm{KS}}$ is the lattice
periodic part of the Kohn-Sham potential, $\mathbf{e}_{\nu}$ and $\omega_{\nu}$ are the phonon eigenvectors and eigenvalues for the $\nu$ mode, and $m_a$ is the mass of the atom.

\begin{figure}[ht]
\centering
\includegraphics[scale=0.55]{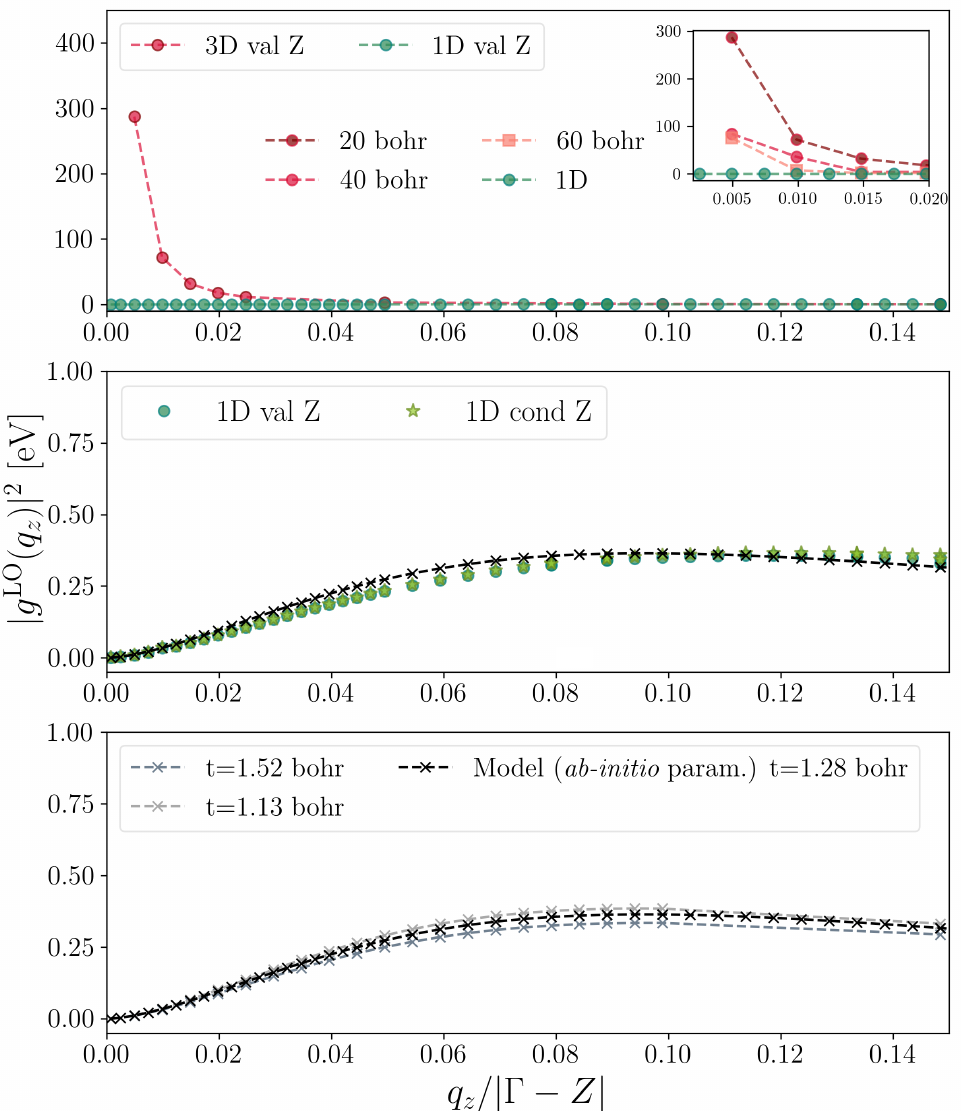}
\caption{Effect of the 1D Coulomb cutoff on EPC matrix elements obtained from DFPT for LO phonons. The analysis is restricted to intraband scattering within the valence or conduction band near $Z$ (although accounting for the degeneracy of the bands as explained).  In the top and central panels, we report in red the results from the standard QE code, 3D PBCs, and in green the ones from 1D-DFPT. In the bottom panel,we overlay the analytical model, summarized in Appendix \ref{models}, onto the 1D-DFPT results for various radii ($t$) used in its parameterization.}
\label{g_comp}
\end{figure}

First-principles results are presented in Fig. \ref{g_comp}, and the 3D PBCs are compared with 1D OBCs (top panel), as was done for phonons.
Very different trends are observed for the small momenta limit of  $|g^{\mathrm{LO}}(\mathbf{q})|^2$.
In fact, in 3D PBCs, the Fr\"{o}hlich interaction diverges as $\mathbf{q}\rightarrow 0$, as expected in 3D bulk materials. 
This occurs no matter the amount of vacuum in the simulation cell (as shown in the inset of the top panel), due to spurious interactions between the periodic images.
Consistent with prior observations, the larger the vacuum, the closer one  asymptotically gets to the isolated case. 
The response from the isolated 1D system is however recovered with the Coulomb cutoff. 
In this case the coupling with LO phonons exhibits a non-monotonic behavior with respect to both 3D and 2D materials. 
In particular, $|g^{\mathrm{LO}} (\mathbf{q}_z)|^2$ goes to 0 at $\Gamma$, reaches a maximum at small momenta and then converges to the 3D $1/q$ behavior for larger momenta.

This trend is found to be in agreement with our analytical model (Appendix \ref{models}), as shown in the center panel of Fig. \ref{g_comp}. 
Note that the analytical $|g^\mathrm{LO}(q_z)|^2$ from the model actually represents the average potential generated by the LO phonons and experienced by the electrons inside the material, where both the electronic and the polarization charge densities are defined via a Heaviside step function with radius $t$.
It does not include the full wavefunction overlap present in DFPT calculations, see Eq. \ref{el-ph_coupling}.
As a result, it loses the information about the dependence on the initial and final electronic states.
Despite this limitation, the comparison between first-principles calculations and the analytical results confirms the qualitative trend of the coupling and the position of the peak at finite but small $q_z$.  
Note that, as shown in the bottom panel of Fig. \ref{g_comp}, this only weakly depends on the choice of the radius $t$. Thus, slightly different parametrizations of $t$ (i.e. different procedures to extract it from the charge density) will lead to a similar result.
In any case, understanding this behavior and being able to predict the peak positions is crucial for several technological applications, including transport.

\subsection{Charge transport}

The developments discussed thus far hold significant relevance for charge transport, which is ubiquitous in various technological applications. 
Specifically, the scattering of electrons with small-momenta phonons is important for doped semiconductors when the Fermi surfaces are small.
The theoretical and computational findings presented here suggest notable modifications in this momentum range.
It seems particularly important to be able to predict the peak position for the coupling, $|g_\mathrm{Fr}(\mathbf{q})|^2$, and its implications on electronic lifetimes, $\tau^{-1}(\epsilon_\mathbf{k})$. 
In the following we provide an approximate estimation of the inverse lifetimes for electrons scattered by LO phonons, evaluated as:
\begin{equation}
\tau_\nu^{-1}(\epsilon_\mathbf{k})=
\frac{2\pi}{\hbar}\sum_{\mathbf{q}}|g_\nu(\mathbf{q)}|^2\delta(\epsilon_{\mathbf{k}+\mathbf{q}}-\epsilon_\mathbf{k}\mp \hbar \omega)
 \left\{\begin{array}{lr}N^{-}_\nu(\mathbf{q},T)\\
 N^{+}_\nu(\mathbf{q},T)
\end{array}\right\} 
\label{tau}
\end{equation}
where the delta function enforces the energy conservation, while $N(\mathbf{q})$ is the Bose-Einstein distribution with index -(+) to indicate phonon absorption (emission). 
We investigate the consequences of the peaked behavior exhibited by the Fr\"ohlich coupling.
In fact, the 1D peaked signature of the coupling is echoed in terms of $\tau^{-1}$.
The peak in this case is shaped –-   shifted and varied in width -- by the Bose-Einstein distribution that accounts for the temperature-dependent phonon population. 
The overall structure is ultimately determined by the relevant phonon momenta (associated with phonon absorption/emission transitions) fulfilling the energy conservation enforced by the Fermi golden rule.

Focusing once more on the BN chain, we consider an initial state at the bottom of the conduction band, around the $Z$ point. For a given initial state, there are two relevant q values associated with phonon absorption processes, and the corresponding coupling strength $|g(\mathbf{q})|^2$ enters the electronic lifetimes.
Building upon the earlier findings, we anticipate the analytical model for the coupling to yield accurate results primarily in the long-wavelength limit, as previously discussed. 
This implies that the model is applicable mainly to scattering events involving electronic states near each other in k-space, contingent upon the effective masses of the relevant bands.
Consequently, for flatter bands, transitions with larger $\mathbf{q}$ values become predominant, resulting in diminished predictive power for the model. 
According to Eq. \ref{tau}, the position of the peak is primarily determined by two factors: the LO phonon frequencies and the curvature of the band near the selected $\mathbf{k}_i$. 
These two factors govern the phonon momenta relevant for the scattering of electrons.
The relevant momenta can be possibly close to the peak of $g$, maximizing the scattering probability, or they can be more or less distant.
This will shape the overall trend of the electronic lifetimes across the bottom of the conduction band.

\begin{figure}[h!]
\centering
\includegraphics[scale=0.47]{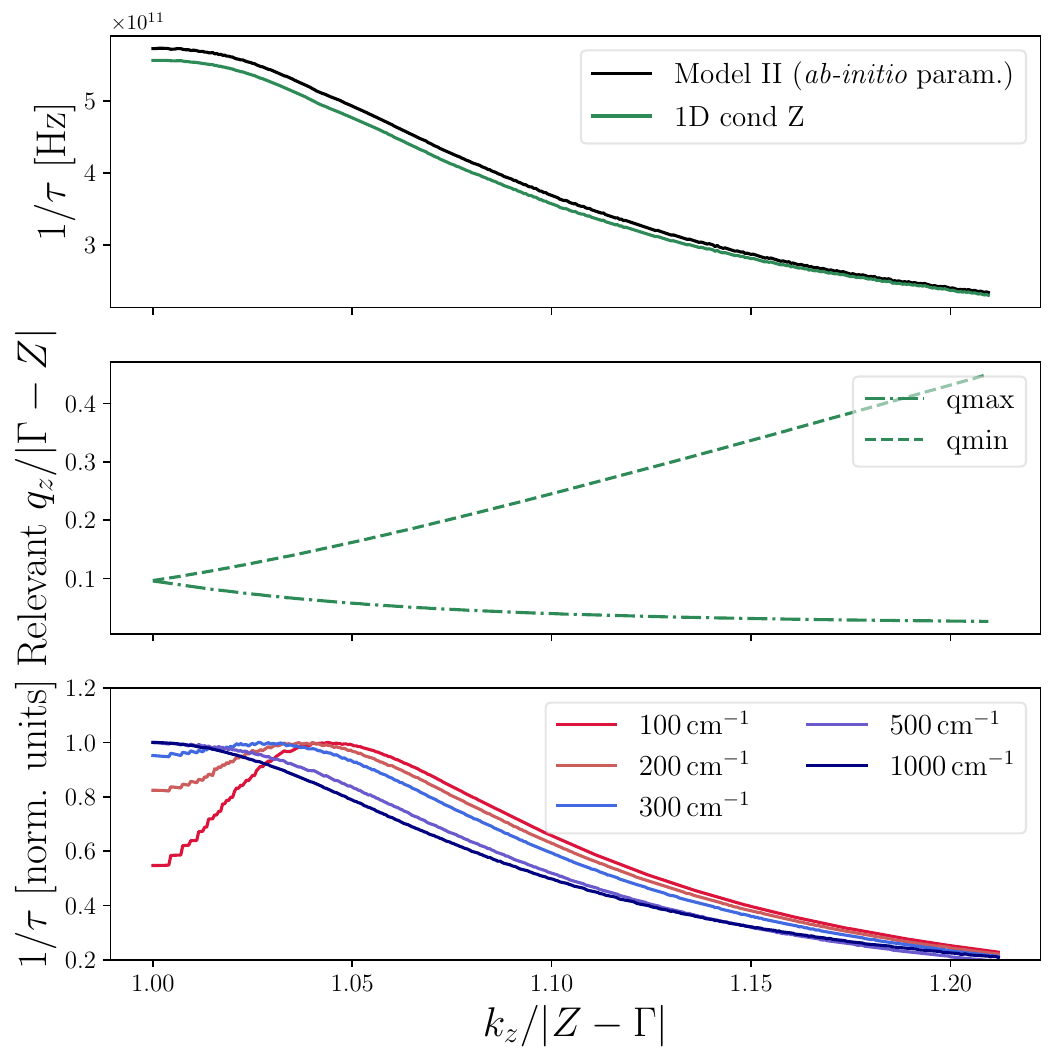}
\caption{1D dimensionality effects on inverse electronic lifetimes due to LO mode scattering in the BN chain.(Top panel) Room temperature inverse lifetimes  for initial states $k_i$ near the bottom of the conduction band, calculated using first-principles phonon and electronic band structures along with the analytical model (black) or 1D-DFPT (green) for the  Fr\"ohlich coupling. (Center panel) The q-points associated to the electronic transitions within the initial k-point range depicted in the top panel. (Bottom) Inverse electronic lifetimes, similar to the top panel, but with the substitution of physical LO frequencies by constant artificial values $\omega$ to tune the range of phonon momenta involved in the transitions.}
    \label{el_lifetimes}
\end{figure}

We present our findings in the three panels of Fig. \ref{el_lifetimes}.
In the top panel, we show the lifetimes obtained through Eq. \ref{tau} for a range of electronic states near the bottom of the conduction band of the BN chain. 
Here, we use first-principles electron and phonon band structures, along with the coupling strength derived analytically and discussed in Appendix \ref{models} and the one obtained via explicit 1D-DFPT; the two are in agreement. In the following we use the analytical results to further comment on the qualitative trends for the lifetimes, clarifying the 1D peculiarities.
Moving to the center panel, we plot the corresponding phonon momenta relevant to each $\mathbf{k}_i$ considered earlier; that is the q-points corresponding to the electronic transition fulfilling the Fermi golden rule in Eq. \ref{tau}.
Notably, the maximum in the inverse lifetimes, that is the strongest scattering, precisely corresponds to the initial electronic states positioned at the bottom of the conduction band (approximately at the $Z$ point). 
This alignment of relevant $q$ values with the peak  position of $|g(\mathbf{q})|^2\times N(\mathbf{q})$ is specific to the conduction band curvature and LO frequency of the BN chain.
A different curvature of the conduction band would change the momenta that satisfy energy conservation.
Alternatively, one could consider different phonon energies.
In the bottom panel, we demonstrate how manipulating the available phonon energy for electronic transitions allows for tuning the relevant $q$ values for each initial $k$-state, consequently shifting the position of the peak.
Smaller phonon energies lead to lower associated phonon momenta for the transitions, progressively shifting the peak further away from the bottom of the conduction band.
Note that the curves have been normalized to emphasize the influence of tuning the available phonon energies on the peak position. Otherwise, large differences in magnitude of $\tau^{-1}$ are observed due to the varying phonon populations, which notably increases for low-energy phonons. This plot is a conceptual representation and does not depict a physical scenario.

\begin{figure}[h!]
\centering
\includegraphics[scale=0.47]{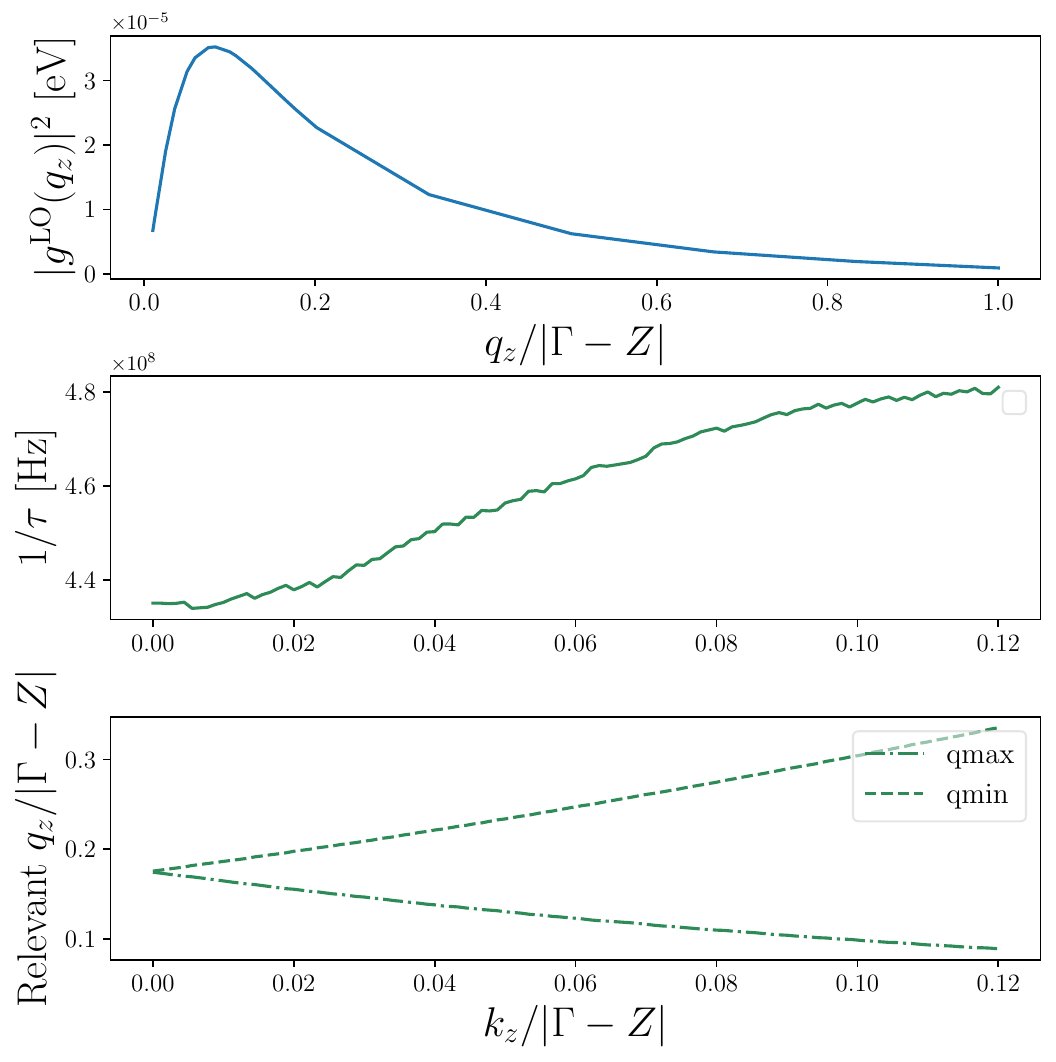}
\caption{1D dimensionality effects on  Fr\"ohlich coupling and  associated inverse electronic lifetimes in the (4,4) BN nanotube. (Top panel) Electron-phonon coupling as a function of phonon momenta obtained from the analytical model. (Central panel) Room temperature results for the inverse lifetimes in the proximity of the bottom of the conduction bands; these are based on first-principles phonon and electronic band structures and the analytic model for the Fr\"ohlich coupling.(Bottom panel) Relevant q-points (two values satisfying energy conservation for each initial electronic state) across the region k-points spanned in panel (central panel).}
    \label{el_lifetimes_NNT}
\end{figure}

To further support our analysis, we use the analytical developments outlined in Appendix B  to qualitatively show the results for the BN nanotube. 
These are shown in Fig. \ref{el_lifetimes_NNT}.
In this case, compared to the chain, the conduction band near its minimum $\Gamma$ is flatter while phonon energies are comparable, and the peak predicted by the model for the coupling happens at relatively larger, but still small, phonon momenta (top panel).
As a result, assuming that the model captures well enough the peak position, the peak (center panel) in inverse lifetimes will happen far from the band extrema, specifically for initial states at $\approx 12 \%$ of the BZ, as shown in the bottom panel. 

 \section{Conclusions}\label{conclusions}

In summary, we introduce a novel DFT and DFPT framework to comprehensively simulate ground-state and, most importantly, linear-response properties of 1D systems from first principles. This achievement is made possible by implementing the 1D Coulomb cutoff in the QE distribution \cite{giannozzi2009quantum,giannozzi2017advanced}; a reciprocal-space technique designed to rectify spurious interactions stemming from periodic images within periodic-boundary conditions.  This restores the proper 1D open-boundary conditions, allowing the intrinsic response of the isolated 1D system to unfold. 
This implementation involves modifying the relevant potentials, enabling the computation of  energies, forces, stresses, and, most notably, phonons and electron-phonon properties.
We then apply our developments to a portfolio of realistic materials that are electrically neutral  with no net spontaneous polarization.
Among the  physical properties affected, we focus extensively on the major example of polar-optical phonons, their dispersion relations and their coupling to electrons,  revealing their strong sensitivity to the dimensionality of the materials.
To complement our DFPT results, we also present an analytical model capable of accurately describing long-wavelength polar optical phonons—those precisely probed by IR and Raman spectroscopies — as well as semi-quantitatively capturing the Fr\"ohlich coupling in 1D materials.
We discuss the characteristic softening of the long-wavelength limit of the LO phonon dispersion curves, as previously observed in a recent paper \cite{rivano2022infrared}.
Moreover, we unveil a novel exotic and non-monotonic behavior, occurring in the same q-limit, regarding the coupling of these phonons to electrons. 
Remarkably, we highlight how this peaked behavior observed in the coupling plays a crucial role in transport applications, resulting in strong scattering for specific initial electronic states. 
These exciting results emphasize how dimensionality provides unparalleled opportunities to tailor material properties. Specifically, we propose that engineering transport properties becomes achievable by strategically tuning the phonon frequencies of the LO modes and/or the curvature of electronic bands.
These results hold profound implications for various practical applications, impacting not only lifetimes and mobilities but also bandgap renormalization and superconducting gaps.
Importantly, all these  results can only be achieved by applying the 1D cutoff, restoring the true physical response and associated signatures.
On top of these novel physical insights, the analytical model also provides us with the fitting contribution to the dynamical matrix coming from the polarity-induced LR interactions. 
This finally allows to smoothly interpolate polar phonons in one-dimensional systems.
In conclusion, our work unlocks the accurate computation of linear-response in 1D systems, deepens our understanding of the dimensional transitions, and sets the stage for similar advancements in the fields of charge transport, optical coupling, and polaritronics.

\section{Aknowledgments}
We acknowledge fundings from the Swiss National Science Foundation (SNSF -- project number 200021-143636) through  the  MARVEL  NCCR and the computational  support  from the  Swiss  National  Supercomputing  Centre  CSCS under project ID mr24.\\

\appendix
\section{Computational details}
First-principles calculations of structural and vibrational properties are performed by combining DFT and DFPT as implemented within the Quantum ESPRESSO distribution \cite{RevModPhys.73.515, giannozzi2009quantum,giannozzi2017advanced} (3D PBCs) and in our modified version with newly implemented 1D periodic-boundary conditions (1D OBCs).
This includes the 1D Coulomb cutoff and a modified phonon Fourier-interpolation based on the analytical model from Reference \cite{rivano2022infrared}. 
The modification of the standard code to include 1D open-boundary conditions will be available at \url{https://github.com/normarivano}. Its release in Quantum ESPRESSO is anticipated, pending the successful integration into the official branch.
We use the Perdew-Burke-Ernzerhof (PBE) exchange-correlation functional\cite{perdew1996generalized}  for all materials and pseudopotentials from the Standard Solid-State Pseudopotentials (SSSP) precision library (version 1.1)  \cite{prandini2018precision}, and the kinetic energy and charge density energy cutoffs have been selected accordingly: 110 and 440 Rydbergs for the chain, 80 and 440 Rydbergs for nanotubes, and 90 and 720 Rydbergs for the GaAs nanowire.
The only exception is bulk wurtzite GaAs, for which we use norm-conserving pseudopotentials within the local density approximation from the original QE PP Library \cite{qepp}  and a kinetic energy cutoff of 80 Ry. 
We treated all the materials under study as non-magnetic insulators (i.e., fixed occupations), and a fine electron-momenta distance of approximately 0.2 $\mathrm{\AA}^{-1}$ has been used to sample the Brillouin Zone.
The convergence of all the relevant parameters has been performed aiming for an accuracy in the final phonon frequencies of a few  $\mathrm{cm}^{-1}$.\\

\section{Analytical models}\label{models}

In Reference \cite{rivano2022infrared}, we addressed an electrostatic problem that led to the development of a comprehensive analytical expression for the electric field generated by polar-optical phonons. This achievement allowed us to derive an analytical formula for the frequencies of these phonons as a function of their momenta and the radius of the material. 
In that work, we described the 1D system as a charge distribution periodic along the $z$-axis and homogeneous in the radial direction within an effective radius $t$, with vacuum outside.
Within the dipolar approximation, the atomic displacement pattern $\mathbf{u}_{\nu}^a$ associated with a phonon $\nu$ of momentum $\mathbf{q} = q_z \mathbf{\hat{z}}$ induces a polarization density $\mathbf{P}(q_z) = \frac{e^2}{L}\sum{a} Z_a \cdot \mathbf{u}^a_{\nu}(q_z)$, where $e$ represents the unit charge, $L$ is the unit-cell length, and $Z_a$ is the BECs tensor for each atom $a$ within the unit cell.
We then solved the Poisson equation associated with this induced polarization
\begin{equation}
\mathbf{\nabla} \cdot[\epsilon \mathbf\nabla{V}_{\mathrm{Fr}}(\mathbf{r})]=-4\pi\mathbf{\nabla} \cdot\mathbf{P}(\mathbf{r}) \,,
\label{poisson}
\end{equation}
by exploiting the periodicity and symmetry of the 1D system, while applying the appropriate electrostatic boundary conditions.
This derivation resulted in a crucial analytical outcome, that is the average interaction potential between these phonons and the electrons. 
This is what we call here  Fr\"ohlich potential and in 1D has the following form:
\begin{widetext}
\begin{equation} 
V_{\mathrm{Fr}}(q_z) =\frac{4e^2}{\epsilon_{\perp}t^2Lq_z}\sum_a \frac{  Z^a \cdot \mathbf{e}^a_{LO} (q_z)} {\sqrt{2M_a\omega_{ q_z LO}}} \Bigl[ 1- \underbrace{2I_1 (|q_z| t) K_1 (|q_z| t) \\\Bigl(1 -  \frac{2 \epsilon_\mathrm{1D} \sqrt\pi q_z t I_1(|q_z| t) K_0(|q_z| t) -G^{2 2}_{2 4}(|q_z|^2 t^2)}{2\sqrt\pi q_z t(\epsilon_\mathrm{1D} I_1(|q_z| t)K_0(|q_z| t)+I_0(|q_z| t)K_1(|q_z| t))}\Bigr)}_{\Delta_{\mathrm{1D}}(q_z,t)}\Bigr]\,, 
\label{gI_II}
\end{equation}
\end{widetext}
where $I_n(x)$, $K_n(x)$  are  the $n^{th}$-order modified cylindrical Bessel functions,  $G_{p q}^{m n}\left(
\begin{array}{c}
a_1,...,a_p\\
b_1,...,b_q
\end{array}\middle\vert x\right)$
is the Meijer G-function, and we assumed a diagonal and isotropic dielectric tensor $\epsilon_\infty$,  i.e., $\epsilon^m\rightarrow\epsilon_{\mathrm{1D}}\mathbb{I})$ (see Ref. \cite{rivano2022infrared} for more details).
The term $\Delta_{\mathrm{1D}}$ carries a clear dimensionality signature.
Specifically, for ($q_zt \to \infty$), $\Delta_{\mathrm{1D}}$ approaches zero, leading to the well-known bulk 3D limit. In this limit, the potential converges to the established prefactor described by the Voigl model \cite{vogl1976microscopic}. Conversely, in the opposite limit, as $q_zt$ tends to zero, $\Delta_{\mathrm{1D}}$ displays a unique 1D asymptotic behavior.
In addition, from this potential, we derived the long-wavelength dispersion relation for polar-optical phonons.
Their dispersion relations  can be recast in the form
\begin{equation}\begin{split}
\omega_{\mathrm{LO}}
=\sqrt{\omega^2_{0}+\frac{ 4 \pi e^2}{\epsilon^{m}_{i} \Omega} \Bigl(\sum_a\frac{ Z_a\cdot\mathbf e^a_{\mathrm{LO}} }{ \sqrt{M_a }}\Bigr)^2 \Bigl[1- \Delta_{\mathrm{1D}}(\mathbf{q},t)\Bigr]}\,,
    \label{omega_1d}
    \end{split}
\end{equation}
where $\omega_{0}$ is the reference value for the LO branch in the absence of any additional contribution from polarity (which can be or not equal to the TO depending on dimensionality and symmetry considerations).

The analytical results displayed in this study rely on
first-principles parameters obtained independently through DFT and DFPT  calculations under 1D open-boundary conditions. In particular, Eqs. \ref{gI_II} and \ref{omega_1d} involve various physical quantities directly derived from such calculations, with the exception of the effective radius $t$ and the in-chain component of the 1D dielectric tensor, i.e., $\epsilon_{\mathrm{1D}}$.
For more details on how these two parameters are obtained, see Sec. \ref{interpol_param} of this paper and the Supplementary Information of Ref. \cite{rivano2022infrared}. Here, for the sake of completeness, we report the radius (in bohr) estimated for each  materials presented: 1.70 for the BN atomic chain, 10.15 for the (4,4) BN nanotubes, add 10.41 for the small GaAs nanowire (24 atoms per unit cell including the hydrogens to saturate the dangling bonds).

\bibliographystyle{plain}

\bibliography{main}
\end{document}